\documentclass{article}
\usepackage{spconf,amsmath,graphicx}
\usepackage{booktabs}
\usepackage{float}

\title{CROSS-LINGUAL MULTI-SPEAKER TEXT-TO-SPEECH SYNTHESIS FOR VOICE CLONING WITHOUT USING PARALLEL CORPUS FOR UNSEEN SPEAKERS}

\name{Zhaoyu Liu {\em and} Brian Mak}
\address{The Hong Kong University of Science and Technology\\
	Department of Computer Science and Engineering\\
	\{zliuar,mak\}@cse.ust.hk}

\graphicspath{{Fig/}}

\begin{document}
\ninept
\maketitle

\begin{abstract}
  We investigate a novel cross-lingual multi-speaker text-to-speech synthesis approach for generating high-quality native or accented speech for native/foreign seen/unseen speakers in English and Mandarin. The system consists of three separately trained components: an x-vector speaker encoder, a Tacotron-based synthesizer and a WaveNet vocoder. It is conditioned on 3 kinds of embeddings: (1) speaker embedding so that the system can be trained with speech from many speakers will little data from each speaker; (2) language embedding with shared phoneme inputs; (3) stress and tone embedding which improves naturalness of synthesized speech, especially for a tonal language like Mandarin.  By adjusting the various embeddings, MOS results show that our method can generate high-quality natural and intelligible native speech for native/foreign seen/unseen speakers. Intelligibility and naturalness of accented speech is low as expected. Speaker similarity is good for native speech from native speakers. Interestingly, speaker similarity is also good for accented speech from foreign speakers. We also find that normalizing speaker embedding x-vectors by L2-norm normalization or whitening improves output quality a lot in many cases, and the WaveNet performance seems to be language-independent: our WaveNet is trained with Cantonese speech and can be used to generate Mandarin and English speech very well.
\end{abstract}

\begin{keywords}
cross-lingual, multi-speaker, text-to-speech, speaker embedding, language embedding
\end{keywords}

\section{Introduction}
\label{sec:intro}

In traditional text-to-speech (TTS) synthesis methods \cite{tts:book.dutoit}, many system components such as the grapheme-to-phoneme model, phoneme duration model, segmentation model, fundamental frequency estimation model and synthesis model are trained separately, and they require expert domain knowledge to produce high-quality synthesized speech. With the advance of deep learning, they are gradually replaced by neural models. 
Deep Voice \cite{deepvoice} presents a neural TTS system which replaces each separate system component with a neural net-based model. The system can synthesize intelligible speech in real time or much faster than real time. In contrast,  Char2wav \cite{char2wav.iclr2017} and Tacotron  \cite{tacotron.is2017} and its improved version Tacotron2 \cite{tacotron2.icassp2018} resort to a totally end-to-end neural model\footnote{Actually ``end-to-end'' here only means that both Char2Wav and Tacotron generate vocoder features, not speech audios, from some representation of input texts.} that use an attention mechanism to convert a sequence of text directly to its corresponding sequence of vocoder features, from which speech audios may be generated using a vocoder. Char2Wav generates WORLD features \cite{world-vocoder.2016} and uses SampleRNN \cite{samplernn} to generate speech, while Tacotron/Tacotron2 generates linear/mel spectrograms and uses the Griffin-Lim (GL) \cite{GriffinLim} and WaveNet \cite{wavenet} vocoder, respectively.  Tacotron 2 can synthesize natural speech comparable to genuine human speech.

Single-speaker neural TTS systems can be readily extended to support multiple speakers’ voices. \cite{MultiSpeakerDNN.2015} takes the multi-task learning approach and duplicates the output layer for each of its training speakers so that each speaker is trained with its own speaker-dependent output layer while sharing other hidden layers in the model. Obviously, the model parameters in its output layer grow linearly with the number of training speakers. Multi-speaker Tacotron \cite{MultiSpeakerTacotron2} is introduced by conditioning Tacotron 2's model on pre-trained d-vector speaker embeddings so that new speakers can be enrolled with a few seconds of speech. Similarly, Deep Voice 2 \cite{DeepVoice2} and Deep Voice 3 \cite{DeepVoice3} extends Deep Voice to multi-speaker TTS. Unlike Tacotron 2, Deep Voice 2 and 3 condition each layer of the model with speaker embeddings which is jointly trained with the rest of the TTS system. For example, Deep Voice 3 claims to support 2400 voices. However, enrollment of new speakers in \cite{DeepVoice2} and \cite{DeepVoice3} will require additional training. VoiceLoop \cite{VoiceLoop} uses a fixed-size memory buffer to accommodate speaker-dependent phonological information and facilitates multi-speaker synthesis by buffer shifts. New speaker embeddings can be trained by an optimization procedure while fixing the other model parameters. Neural Voice cloning \cite{NeuralVoiceCloning} introduces a similar speaker adaptation method where both model parameters and speaker embeddings are fine-tuned with data from the new speaker.
Multi-lingual TTS further extends multi-speaker TTS to support synthesis in more than one language.
For example, \cite{chen2019cross} introduces a cross-lingual TTS system in English and Mandarin trained with IPA without language embedding. It succeeds in synthesizing speech in two languages, however, it can only synthesize native speech but not accented speech. It uses the GL vocoder (instead of WaveNet or other neural-based high fidelity vocoders) resulting in synthesized speech of lower quality.

\begin{figure*}[tbh!]
  \centerline{\includegraphics[width=0.8\textwidth]{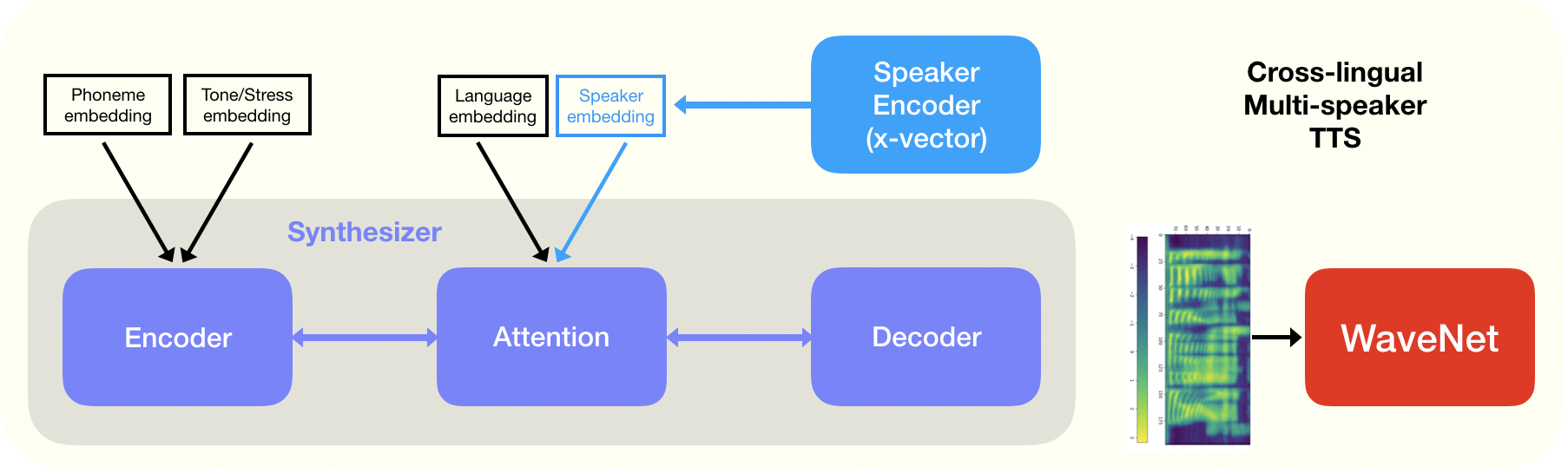}}
  \caption{Multi-lingual multi-speaker TTS system using speaker embedding, language embedding and tone/stress embedding.}
  \label{fig:tts}
\end{figure*}

In this paper, we investigate an approach to synthesize high-quality speech\footnote{Some audios synthesized with different settings are available at https://zliuar.github.io/CLMSTTS/.} that sounds like native speech or accented speech spoken by native or foreign speakers of a language, who may be seen or unseen during training of the TTS system. 
While we are preparing for this submission, we notice a new non-refereed publication from Google \cite{MultilingualTacotron2} on arXiv recently which shares many ideas in our system. Nonetheless, there are the following notable differences:
\begin{itemize}
    \item Most importantly, our results are reproducible as we use publicly available training corpora: English from LibriSpeech and Mandarin from SurfingTech, while the system in \cite{MultilingualTacotron2} is trained on proprietary data.
    \item \cite{MultilingualTacotron2} aims at synthesizing speech with training speakers' voices; thus, their training data consists of few speakers (some are professional voice actors) but each has tens of hours of speech. On the contrary, we train our system on hundreds of speakers with less than 25 minutes of speech from each speaker. We believe our system is more generalizable to new speakers and we report results on unseen speakers while \cite{MultilingualTacotron2} does not.
    \item Both our system and theirs employ shared phonemes for inputs and speaker embeddings, language embedding, and stress and tone embeddings. However, we use the state-of-the-art x-vector for speaker embedding while theirs is d-vector. We expect our synthesized speech will be better in terms of speaker similarity, especially for unseen test speaker.
    \item Our model is simpler with no residual encoding nor adversarial training. Instead, we investigate on the effect of various normalization methods on the speaker embedding vectors for enhancing the intelligibility, naturalness and speaker similarity of the synthesized speech.
    \item We also investigate the effect of training the WaveNet vocoder with a third language (Cantonese) to synthesize speech of the intended languages (English and Mandarin) in the system.
\end{itemize}

The rest of the paper is organized as follows. Section 2 describes the details of our model architecture. Section 3 introduces experiments conducted and their corresponding settings. It also shows the MOS results on the naturalness, intelligibility and speaker similarity of the synthesized speech from our system. Finally, Section 4 gives our concluding remarks.

\section{MODEL STRUCTURE}
\label{sec:model}

Fig. \ref{fig:tts} shows our multi-lingual multi-speaker TTS system.

\subsection{Input Representation: Phoneme, Tone, Stress Embeddings}

Instead of character embedding in \cite{char2wav.iclr2017, tacotron.is2017, tacotron2.icassp2018}, we use phoneme embedding which has been proved to generate more natural speech. A shared phoneme set is created by mapping all pinyin to ARPABET phonemes, with the exceptions of pinyin phonemes ‘j’, ‘q’ and ‘x’ which are treated as distinct phonemes as no good ARPABET mappings are found.
We separately represent tone and stress as 7-D one-hot embeddings concatenated to phoneme embedding. No tone and no stress are sharing the same representation. 

\subsection{Speaker Encoder}

The speaker encoder use x-vectors described in \cite{x-vector} as speaker embeddings. X-vectors are derived from a TDNN-based speaker discriminative model that maps variable-length utterances to fixed-dimensional speaker embeddings. X-vectors are trained using Kaldi \cite{Kaldi} with the standard experiment settings. We extract x-vectors from the output of 6th hidden layer in the TDNN.

\subsection{Mel-spectrogram Synthesizer}

The mel-spectrogram synthesizer is implemented based on multi-speaker Tacotron 2 \cite{MultiSpeakerTacotron2} with the same architecture and parameter settings. 
We concatenate speaker embeddings and language embedding to the encoder context output as in \cite{MultiSpeakerTacotron2}.
In our pre-experiments, we have tried linear addition instead of concatenation. We have also tried to input language embedding before or after the convolutional layers in the text encoder by either addition or concatenation. Both do not lead to any significant improvement.

\subsection{WaveNet}

WaveNet \cite{wavenet} is an auto-regressive sample-by-sample raw audio synthesizer.
We construct the WaveNet with 30 layer of dilated causal convolutions. 
In our work, we train the WaveNet with 8-bit mu-law quantization\footnote{Training a WaveNet with 8-bit mu-law outputs allows much faster convergence and the output quality is still very good.} using the CUSENT Cantonese corpus and demonstrate that it can still synthesize high-quality and natural speech in both English and Mandarin.

\begin{figure*}[htb]
 \begin{minipage}{0.33\textwidth}
     \centering
         \includegraphics[width=1\linewidth]{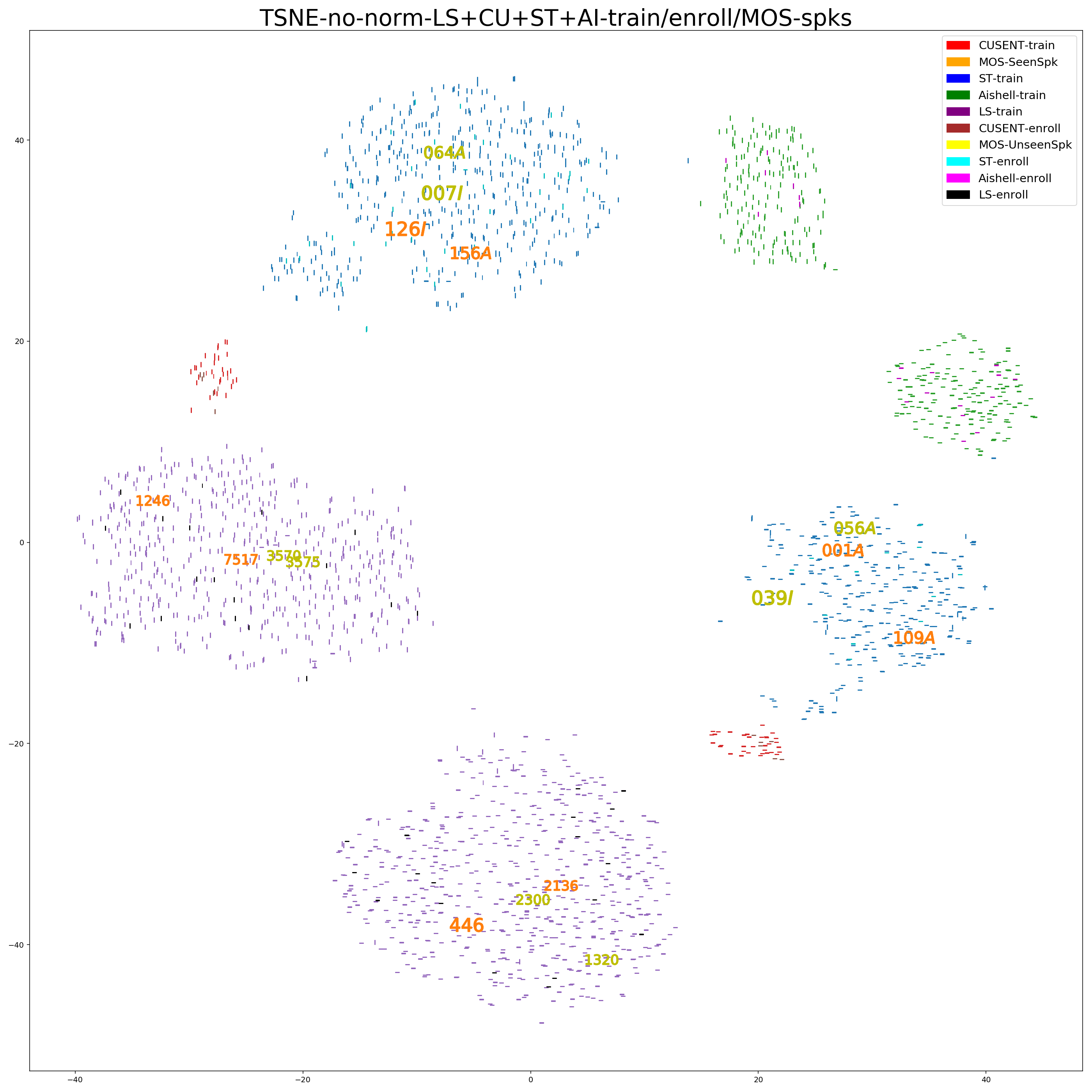}
     \caption{No-norm}\label{Fig:2a}
  \end{minipage}\hfill
 \begin{minipage}{0.33\textwidth}
     \centering
     \includegraphics[width=1\linewidth]{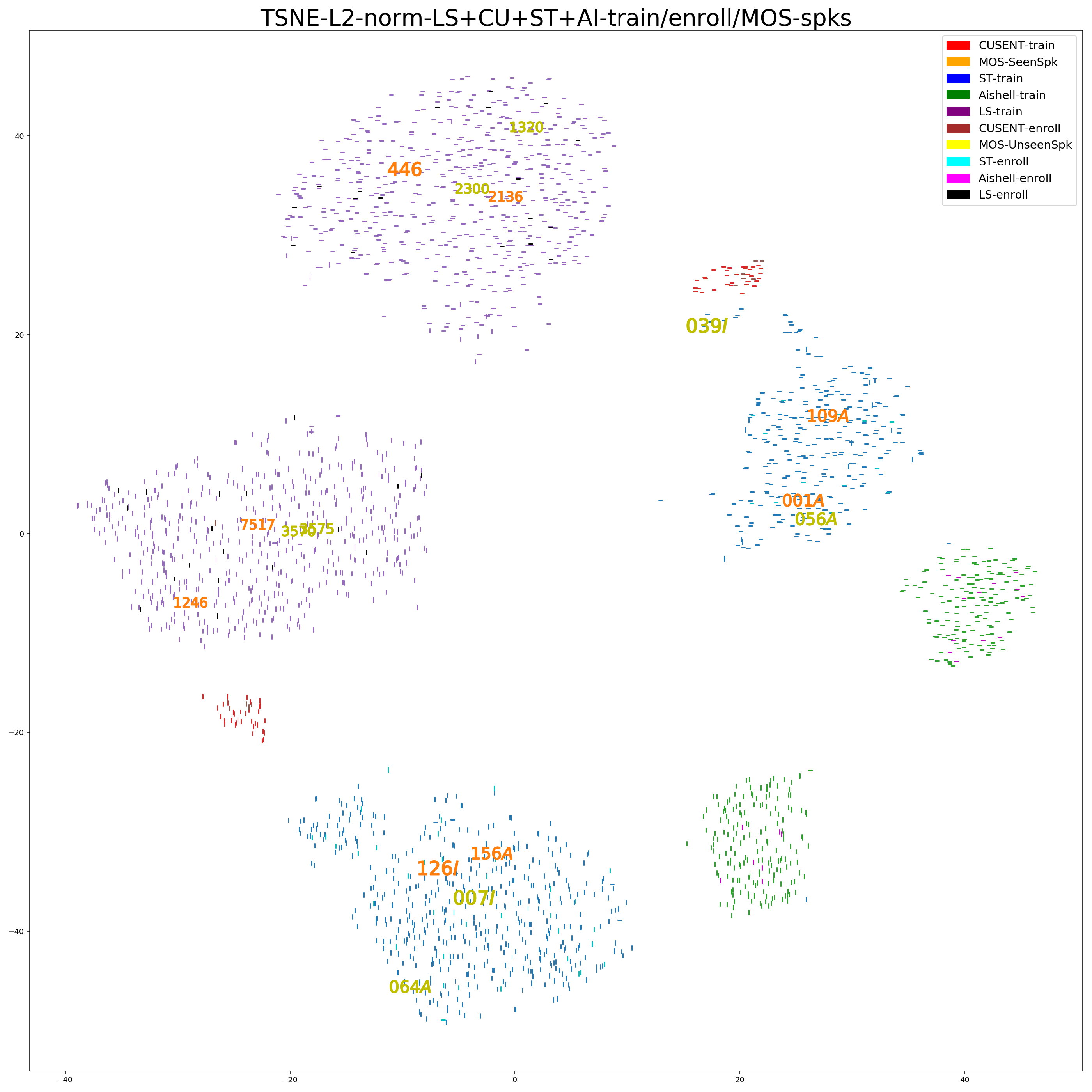}
     \caption{L2-norm}\label{Fig:2b}
  \end{minipage}\hfill
  \begin{minipage}{0.33\textwidth}
     \centering
     \includegraphics[width=1\linewidth]{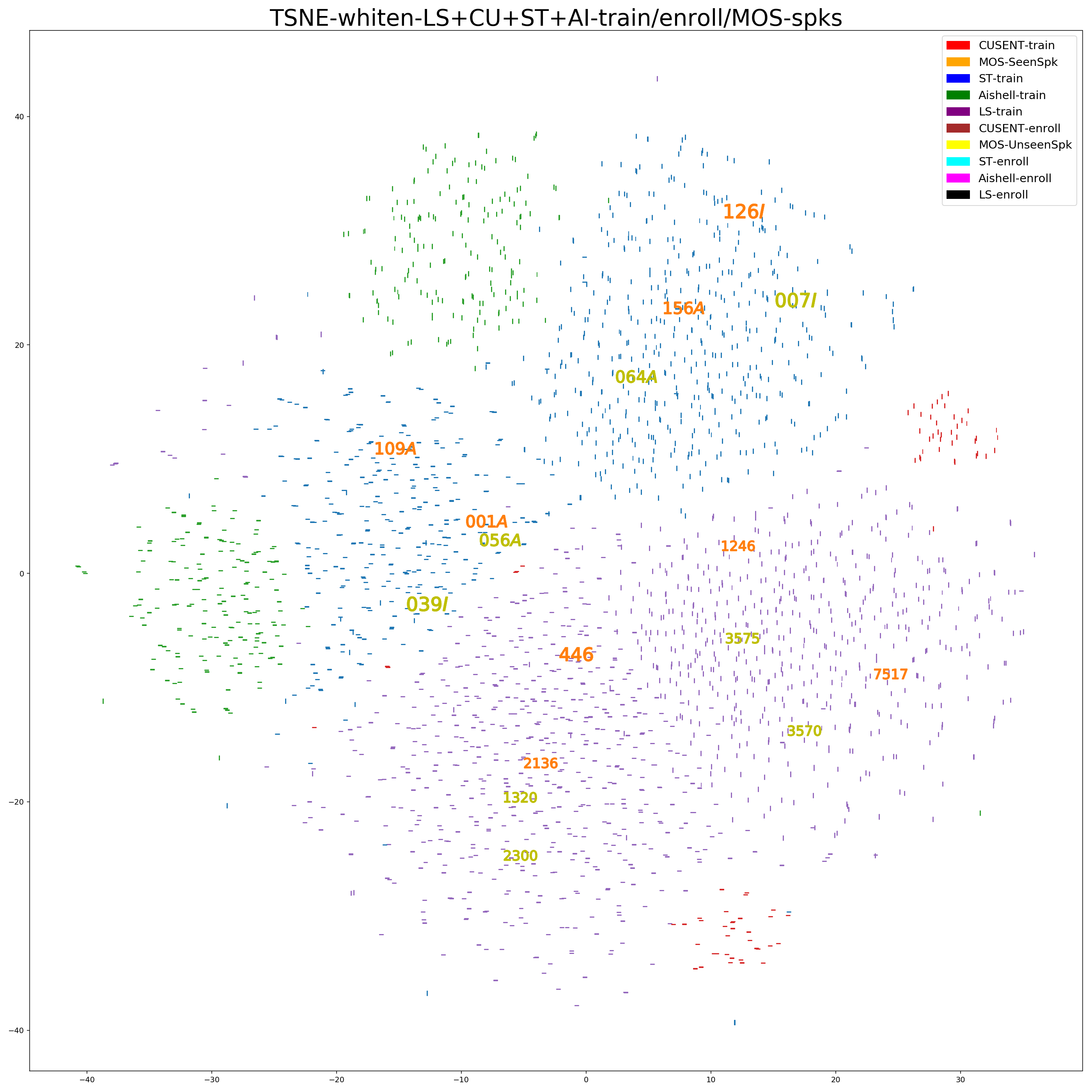}
     \caption{Whitening}\label{Fig:2c}
  \end{minipage}
\end{figure*}

\section{Experiments \& Results}
\label{sec:pagestyle}

We train our model on 4 open-source datasets in three languages: (1) The ``clean'' set in Librispeech (LS) \cite{Librispeech} consists of 1172 English speakers, each with 25 minutes of speech; (2) SurfingTech (ST) \cite{surfingtech} is a Mandarin corpus which has 855 speakers, and 10 minutes of speech per speaker. We further randomly select 800 speakers' data for training; (3) CUSENT (CU) \cite{Cusent} is a Cantonese corpus of 20 hours of speeh from 68 training speakers; (4) Aishell (AI) \cite{aishell} is a Mandarin corpus consisting of 150 hours of speech from 340 training speakers. We use all data to train the x-vector speaker encoder, and use Librispeech and SurfingTech data to train the synthesizer. CUSENT is used to train WaveNet only. Librispeech data are segmented by forced alignment to shorter audios (2s-12s) upon silences that are longer than 0.3s. Block Threshold denoising \cite{BlockThresholding} is used to denoise the segmented Librispeech. We use Google Translate to transform text transcriptions to pinyin which are then mapped to ARPABET phonemes. Forced alignment is performed on SurfingTech to label significant short pauses in its speeches. 

\subsection{Speaker Verification Evaluation}
\label{ssec:subhead}

We conduct objective speaker verification (SV) evaluations on x-vector speaker embeddings with increasing number of training speakers. Enrolment utterances are 3 minutes long and test utterances vary from 5--12s. We first test 400-D i-vectors, 64/128/512-D x-vectors on Librispeech SV, and the results are shown in
Table \ref{tab:LSEER}. In our experiments, even though the EERs differs from 1 to 3.25 for various models, we found the TTS systems using the i-vectors or 128-D x-vectors can generate better speech with very similar quality. In contrast, although 64-D x-vectors give the best SV EER, they produce audios of poorer quality in our TTS system. It seems that embeddings that give better SV EER is no guarantee for better synthesized audios.
At the end, we choose the 128-D x-vectors for our speaker embeddings.
We further investigate the x-vector embedding with more training speakers. Table \ref{tab:MLEER} shows SV EER on a separate test set for systems trained with increasing number of speakers from various corpora. Results show adding more speakers from other datasets does not further reduce the SV EER. We visualize these x-vectors using t-SNE in Fig. \ref{Fig:2c}. It shows that speakers from different datasets form their own clusters without interfering each other. We perform L2-norm and whitening normalization to the x-vectors.
L2-norm is supposed to preserve cosine similarity and whitening will remove correlations among the dimensions. Fig. \ref{Fig:2b} and \ref{Fig:2c} shows the normalized x-vectors.

\begin{table}[tbh]
\centering
\begin{tabular}{|c|c|c|c|c|}
\hline
System & Dim & Train set & Speakers & SV-EER \\ \hline \hline
i-vector & 400 & LS & 1172 & 3.25\\ \hline
x-vector & 64 & LS & 1172 & 1.00  \\ \hline
x-vector & 128 & LS & 1172 & 1.50 \\ \hline
x-vector & 512 & LS & 1172 & 1.25 \\ \hline
\end{tabular}
\caption{Librispeech SV EER (\%) for various speaker embeddings. }
\label{tab:LSEER}

\vspace{1.5em}

\centering
\begin{tabular}{|c|c|c|c|c|c|}
\hline
Train set & Speakers & LS & CU & ST & AI \\ \hline \hline
LS, CU & 1240 & 0.75 & 0 & -- & -- \\ \hline
LS, ST, AI & 2312 & 0.75 & -- & 0 & 0.5 \\ \hline
LS, CU, ST, AI & 2380 & 0.75 & 0 & 0 & 0.5 \\ \hline
\end{tabular}
\caption{Effect of increasing number of training speakers on Librispeech SV EER (\%) using 128-D x-vector.}
\label{tab:MLEER}

\vspace{1.5em}

\centering
\setlength\tabcolsep{2pt}
\begin{tabular}{ccccccccc}
\hline
Seen & 446 & 1246 & 2136 & 7517 & 1A & 109A & 126I & 156A \\ \hline
Gender & M & F & M & F & M & M & F & F \\
Dataset & LS & LS & LS & LS & ST & ST & ST & ST \\ \hline
Unseen & 1320 & 2300 & 3570 & 3575 & 7I & 39I & 56A & 64A \\ \hline
Gender & M & M & F & F & F & M & M & F \\
Dataset & LS & LS & LS & LS & ST & ST & ST & ST \\ \hline
\end{tabular}
\caption{MOS test speakers: Gender: Male (M) and Female (M); Dataset: LS and ST}
\label{tab:Testspks}
\end{table}

\subsection{Subjective Evaluation}

We evaluate our TTS system subjectively with mean opinion score (MOS) tests. First, we train an English multi-speaker TTS system as our baseline model with 128-D x-vectors trained on Librispeech. We construct three multi-lingual TTS models with unnormalized, L2-norm normalized and whitened 128-D LS+CU+ST+AI x-vectors. We perform MOS evaluation on the four models on intelligibility (I), naturalness (N) and speaker similarity (S) with Absolute Category Rating Scale from 1--5 with 0.5 increments. Table \ref{tab:Testspks} lists test speakers where English speakers are a subset of test speakers in \cite{MultiSpeakerTacotron2} and Mandarin speakers are randomly selected. Their native/accented utterances are synthesized from 4 VCTK \cite{VCTK} English texts and 4 Aishell Mandarin texts to form an evaluation pool of 192 synthesized utterances for the multi-lingual model and 32 for the baseline model. An evaluation set of 100 utterances is constructed with 16 ground truth utterances, 12 synthesized utterances from the baseline model, and 24 synthesized utterances from each multi-lingual model. Samples are randomly selected from each speech pool with gender-balanced seen/unseen speakers. The 72 synthesized utterances from multi-lingual models consists of generated utterances in foreigner's accented English (FAE), foreigner's native English (FNE) and native English (NE), and the corresponding FAM, FNM and NM (for Mandarin samples), and they are presented in that order lest the raters learn the content from the more intelligible speeches. Ten human raters are recruited from our speech lab, who all speak Mandarin as their mother language and are proficient in English.

Table \ref{tab:GBMOS} shows MOS scores on ground truth speech and synthesized speech from the baseline model on I/N/S. We find that some raters rate synthesized speech more intelligible than ground truth speech. It turns out some of the ground truth utterances from LS and ST are drawn from texts containing words that are not familiar to some raters. Ground truth speech in LS is rated as extremely natural but lower speaker similarity than ST because of emotion and speaker style changes in reading audio books. ST speeches are rated as less natural because of their various accents different from standard Mandarin. Our baseline model obtains a high score in naturalness for both seen and unseen speakers, and speaker similarity between moderately similar and very similar.

Table \ref{tab:UnseenMosI} shows intelligibility MOS on unseen speakers for multi-lingual models. We observe that raters rate FAE and FAM bad/poor. However, they regret when they evaluate the synthesized native speech (which is highly intelligible) and learn the actual texts. FNE and FNM speech with no normalization of the x-vectors are rated bad/poor because the model does not produce clear speech in this case and most of FNM speech do not stop properly. Both L2-norm and whitening normalization result in high-quality synthesized speech for unseen speakers that are rated between good and extremely intelligible.

Table \ref{tab:UnseenMosN} shows naturalness MOS results. We observe whitening helps generate very natural native English and foreign native Mandarin for Librispeech speakers. 
The naturalness of foreign accented speech is bad though L2-norm normalization helps produce better results. Table \ref{tab:SeenMosS} and \ref{tab:UnseenMosS} show the speaker similarity MOS results on seen and unseen speakers, respectively. As expected, the results on seen speakers are better than those on unseen speakers. However, we find a very interesting phenomenon for the case of Mandarin: speaker similarity on FAM from English speakers is better than that on NM from native Mandarin speakers.

In summary, the MOS results on intelligibility, naturalness and speaker similarity of synthesized speech from our model are better for English than Mandarin. There are two plausible explanations: (a) the quality of  Librispeech English speech is higher than that of the SurfingTech Mandarin speech which also come from speakers with various accents; (b) the raters are all native Chinese who may be more critical in assessing the quality of speech of their mother tongue.

\section{Conclusion}
\label{sec:conclusion}

This paper presents a novel multi-lingual multi-speaker TTS approach which performs no worse than its mono-lingual baselines. With x-vector whitening, it can produce very natural native English and Mandarin compared to the ground truth speech with moderate speaker similarity on unseen speakers. The model can successfully clone foreigners' voices to speak another language like a native intelligibly and naturally (FNE and FNM results). It can also generate accented speech for foreign speakers.  We further find that a WaveNet trained on Cantonese can generate high-quality speech in Mandarin and Engish; it seems WaveNet training is language-independent.

\section{Acknowledgements}
The work described in this paper was partially supported by grants from the
Research Grants Council of the Hong Kong Special Administrative Region, China
(Project Nos. HKUST16200118 and HKUST16215816).

\begin{table}[tbh]
\centering
\renewcommand{\arraystretch}{1.1}
\begin{tabular}{cccc}
\toprule
 & Intelligibility & Naturalness & Similarity \\ \midrule
GT-E &4.47 $\pm$ 0.22 &4.90 $\pm$ 0.14 & 4.33 $\pm$ 0.24\\ \cmidrule(r){1-1} 
GT-M &4.68 $\pm$ 0.18 &4.30 $\pm$ 0.24 & 4.86 $\pm$ 0.15 \\ \cmidrule(r){1-1} 
Baseline seen & 4.96 $\pm$ 0.06 & 4.25 $\pm$ 0.26 & 3.82 $\pm$ 0.27\\ \cmidrule(r){1-1} 
Baseline unseen & 4.50 $\pm$ 0.19 &4.11 $\pm$ 0.32 & 3.40 $\pm$ 0.33 \\ \bottomrule
\end{tabular}
\caption{MOS on ground truth speech and synthesized speech from baseline models.}
\label{tab:GBMOS}

\vspace{1em}

\centering
\begin{tabular}{@{}ccccccc@{}}
\toprule
Model\textbackslash{}Cases & FAE & FNE & NE \\ \hline
No-norm &2.08 $\pm$ 0.23 & 3.50 $\pm$ 0.52 & 4.67 $\pm$ 0.18\\ \cline{1-1}
L2-norm & 2.29 $\pm$ 0.35 & 4.92 $\pm$ 0.10 & 4.54 $\pm$ 0.21   \\ \cline{1-1}
Whitening & 1.92 $\pm$ 0.27 & 4.25 $\pm$ 0.24 & 4.92 $\pm$ 0.07\\ \hline
Model\textbackslash{}Cases & FAM & FNM & NM \\ \hline
No-norm &  1.75 $\pm$ 0.23 & 2.83 $\pm$ 0.25 & 4.77 $\pm$ 0.14 \\ \cline{1-1}
L2-norm & 1.88 $\pm$ 0.29 & 4.46 $\pm$ 0.17 & 4.71 $\pm$ 0.15 \\ \cline{1-1}
Whitening &  1.42 $\pm$ 0.19 & 4.08 $\pm$ 0.44 & 4.67 $\pm$ 0.13 \\ \hline
\end{tabular}
\caption{Intelligibility MOS with unseen speakers.}
\label{tab:UnseenMosI}

\vspace{1em}

\centering
\begin{tabular}{@{}ccccccc@{}}
\toprule
Model\textbackslash{}Cases & FAE & FNE & NE \\ \hline
No-norm & 1.88 $\pm$ 0.22 & 3.79 $\pm$ 0.28 & 4.33 $\pm$ 0.22 \\ \cline{1-1}
L2-norm &   1.96 $\pm$ 0.09 & 4.08 $\pm$ 0.19 & 4.08 $\pm$ 0.23 \\ \cline{1-1}
Whitening & 1.47 $\pm$ 0.16 & 3.46 $\pm$ 0.26 & 4.58 $\pm$ 0.19 \\ \hline
Model\textbackslash{}Cases & FAM & FNM & NM \\ \hline
No-norm & 1.88 $\pm$ 0.38 & 1.92 $\pm$ 0.26 & 4.18 $\pm$ 0.23  \\ \cline{1-1}
L2-norm & 2.25 $\pm$ 0.36 & 3.17 $\pm$ 0.48 & 4.21 $\pm$ 0.24 \\ \cline{1-1}
Whitening & 1.75 $\pm$ 0.28 & 3.58 $\pm$ 0.33 & 4.33 $\pm$ 0.20\\ \hline
\end{tabular}
\caption{Naturalness MOS on unseen speakers.}
\label{tab:UnseenMosN}

\vspace{1em}

\centering
\begin{tabular}{@{}ccccccc@{}}
\toprule
Model\textbackslash{}Cases & FAE & FNE & NE \\ \hline
No-norm &  2.62 $\pm$ 0.28 & 1.94 $\pm$ 0.14 & 3.75 $\pm$ 0.38 \\ \cline{1-1}
L2-norm &  2.69 $\pm$ 0.23 & 2.75 $\pm$ 0.29 & 4.19 $\pm$ 0.35 \\ \cline{1-1}
Whitening &  3.36 $\pm$ 0.24 & 3.06 $\pm$ 0.21 & 3.81 $\pm$ 0.25 \\ \hline
Model\textbackslash{}Cases & FAM & FNM & NM \\ \hline
No-norm & 4.06 $\pm$ 0.19 & 1.43 $\pm$ 0.25 & 4.08 $\pm$ 0.20\\ \cline{1-1}
L2-norm & 3.75 $\pm$ 0.18 & 1.75 $\pm$ 0.28 & 3.12 $\pm$ 0.29 \\ \cline{1-1}
Whitening &3.94 $\pm$ 0.16 & 1.38 $\pm$ 0.15 & 3.47 $\pm$ 0.43\\ \hline
\end{tabular}
\caption{Speaker similarity MOS with seen speakers.}
\label{tab:SeenMosS}

\vspace{1em}

\centering
\begin{tabular}{@{}ccccccc@{}}
\toprule
Model\textbackslash{}Cases & FAE & FNE & NE \\ \hline
No-norm &2.69 $\pm$ 0.37 & 2.62 $\pm$ 0.32 & 3.46 $\pm$ 0.30 \\ \cline{1-1}
L2-norm & 2.46 $\pm$ 0.34 & 2.29 $\pm$ 0.22 & 3.37 $\pm$ 0.32 \\ \cline{1-1}
Whitening & 2.19 $\pm$ 0.34 & 2.08 $\pm$ 0.35 & 3.50 $\pm$ 0.23 \\ \hline
Model\textbackslash{}Cases & FAM & FNM & NM \\ \hline
No-norm &3.21 $\pm$ 0.25 & 1.25 $\pm$ 0.14 & 3.32 $\pm$ 0.41\\ \cline{1-1}
L2-norm &  3.75 $\pm$ 0.21 & 2.08 $\pm$ 0.27 & 3.17 $\pm$ 0.29 \\ \cline{1-1}
Whitening & 3.75 $\pm$ 0.14 & 1.67 $\pm$ 0.15 & 3.33 $\pm$ 0.32\\ \hline
\end{tabular}
\caption{Speaker similarity MOS with unseen speakers.}
\label{tab:UnseenMosS}
\end{table}


\clearpage
\bibliographystyle{IEEEtran}
\bibliography{abbrev,refs}

\end{document}